\documentclass[final]{svjour2}
\usepackage{graphicx}
\usepackage{exscale}
\usepackage{rotating}
\usepackage{amssymb}
\usepackage{mathptmx}
\usepackage{bm}
\usepackage[numbers]{natbib}
\usepackage[nofighead,nomarkers]{endfloat}
\makeatletter
\journalname{Journal of low temperature physics}
\bibpunct{}{}{,}{s}{}{,}

\begin{document}

\newcommand{\hdblarrow}{H\makebox[0.9ex][l]{$\downdownarrows$}-}

\title{Spin Path Integrals, Berry phase, and the Quantum Phase Transition in the sub-Ohmic Spin-boson Model}

\author{Stefan Kirchner}

\institute{Max Planck Institute for the Physics of Complex Systems and,\\
Max Planck Institute for Chemical Physics of Solids,\\ 01187 Dresden, Germany\\ 
Tel.:+49(0)351-871-1121\\ Fax::+49(0)351-871-1999\\
\email{kirchner@pks.mpg.de}}

\date{}

%\runningheads{S. Kirchner}{Spin path integrals and the QCP in the spin-Boson model}
\maketitle


\begin{abstract}
The quantum critical properties of the  sub-Ohmic spin-1/2 spin-boson model and of the Bose-Fermi Kondo model have
recently been discussed controversially. The role of the Berry phase in the breakdown of the quantum-to-classical 
mapping of quantum criticality in the
spin-isotropic Bose-Fermi Kondo model has been discussed previously.
In the present article, some of the subtleties underlying  the functional integral representation of
the spin-boson and related models with spin anisotropy are discussed. 
To this end, an introduction to spin coherent states and spin path integrals is presented with a focus on the
spin-boson model.
It is shown that, even for the Ising-anisotropic case as in the spin-boson model, the path integral in the 
continuum limit in the coherent state representation involves a Berry phase term.
As a result, the effective action for the spin degrees of freedom does not assume 
the form of a Ginzburg-Landau-Wilson functional.
The implications of the Berry-phase term for the quantum-critical behavior of the spin-boson model are discussed.
The case of arbitrary spin $S$ is also considered.

\end{abstract}

\section{Introduction}
\label{sec:GLW}
A quantum critical point (QCP) separates thermodynamic phases of matter at zero temperature and occurs when 
the order vanishes continuously as the zero-temperature phase transition is approached from the 
ordered side\cite{Sachdev,Loehneysen.07}.
Quantum criticality has attracted considerable attention in recent years. This is not only due to the fact that
quantum critical fluctuations are important in a broad temperature range around a  QCP but has largely been fueled by
the realization that not all continuous zero-temperature phase transitions may
be described in terms of an order-parameter functional in elevated
dimension\cite{Natphysi_focus,Si.01,Coleman.01,Senthil.04}. 
The clearest experimental indication for such unconventional quantum criticality
has so far come from anti-ferromagnetic heavy fermion metals\cite{Gegenwart.07}. 
The magnetic energy scales in these systems are small enough to tune the
system from an itinerant anti-ferromagnet to a paramagnetic metal
as an external tuning parameter is varied.
In the traditional approach to the QCP in itinerant
anti-ferromagnets,  universal properties are described
in terms of a Ginzburg-Landau-Wilson functional of the order parameter and its
fluctuations resulting in a $\phi^4$-theory in $D+z$ dimensions, where $z=2$ is the dynamical exponent and $D$ is the spatial
dimension of the system\cite{Hertz.76,Millis.93}.  In contrast, a number of
compounds which can be tuned through a zero-temperature transition do not
follow the predictions of such a spin-density wave (SDW) theory\cite{Aronson.95,Schroeder.00,Park.06,Friedemann.09}.
This has led to the understanding that the concomitant destruction of the
Kondo screening of the f-moments may result in an altered critical fluctuation spectrum\cite{Coleman.01,Si.01}.
One scenario, characterized by critical Kondo screening at the
zero-temperature phase transition that is able to explain the experimental findings that are at
variance with the conventional SDW picture has been termed 'local quantum
criticality'\cite{Si.01,Si.03}. 
A microscopic approach to the problem  which preserves the
dynamic competition between Kondo screening and magnetic correlations and
consequently can capture the critical Kondo destruction is given
by the Extended Dynamical Mean Field Theory (EDMFT)\cite{Si.01,Si.05}. Within
EDMFT, the Kondo lattice is mapped onto a quantum impurity model where
a local moment is coupled to a
fermionic and a bosonic bath  and augmented with self-consistency 
requirements\cite{Si.03}. This quantum impurity model is called the Bose-Fermi Kondo model (BFKM).
The BFKM
is, however, not only of interest in the context of the EDMFT approach to quantum critical heavy fermion compounds.
The BFKM
is also the effective low-energy model of  single-electron transistors with ferromagnetic
leads\cite{Kirchner.05a,Kirchner.08c} and of Coulomb-blockaded noisy quantum dots\cite{LeHur.04}.
An interesting and important question is,  under which conditions the QCP in the BFKM
may be unconventional.
After all, the EDMFT approach to the Kondo lattice preserves the dynamic  competition between the 
local Kondo screening at each
f-moment site and the magnetic fluctuations among different f-moments.
\newline
%The local QCP occuring in the selfconsistent version of the Bose-Fermi Kondo model
%is expected to have the same critical properties as the one occuring in the
%Kondo lattice.
So far, this question has been addressed in the spin-isotropic and the easy-axis BFKM and 
the spin-boson model\cite{Zhu.04,Glossop.05,Kirchner.09,Vojta.05,Winter.09,Vojta.09}. 
Based on a dynamical large-N limit, it was shown that the  QCP in the spin-isotropic BFKM is in general not 
described in terms
of a local $\phi^4$-theory\cite{Zhu.04}. This is in contrast to the critical behavior of a classical spin chain  
in  $0+1$ dimension
with a long-ranged interaction that is compatible with the retarded spin-spin interaction of the 
BFKM generated by the coupling to the fermionic and bosonic degrees of freedom\cite{Fisher.72,Suzuki.76}.
As it turns out, the difference between the critical properties of the spin-isotropic BFKM and the classical spin chain 
is not spoiled by 1/N-corrections but can be 
traced back to the presence of a  Berry phase term in 
the spin path integral representation of the quantum problem\cite{Kirchner.08,Kirchner.08e}.
The quantum-to-classical mapping of quantum criticality that links a quantum-critical system to a 
$D+z$-dimensional classical critical system,
therefore does not hold for the spin-isotropic BFKM~\cite{Kirchner.08e}.
\newline
The situation is at present less clear for the  (spin-anisotropic) easy-axis BFKM and the spin-boson problem.
While the two models are expected to belong to the same universality class, 
it is not clear whether the critical theory of the two models is identical to
those of a local $\phi^4$-theory\cite{Kirchner.09,Winter.09}.
Numerical Renormalization Group (NRG) calculations point to a critical theory that is different from a 
local $\phi^4$-theory~\cite{Vojta.05,Glossop.05,Glossop.07b,Kirchner.09}. 
The validity of such  NRG calculations for critical systems involving bosons
has however been called into question~\cite{Winter.09,Vojta.09,Vojta.10}. 
A recent continuous time Monte Carlo study~\cite{Winter.09} addresses the
dissipative spin-boson model based on a Monte Carlo algorithm
that explicitly takes the limit of vanishing lattice constant of a one-dimensional
classical Ising chain~\cite{Rieger.99}.
In reference~\cite{Kirchner.09}, it was pointed out that the continuum limit of the classical Monte Carlo misses
the effects associated with properly regularizing the spin-flip terms. These topological excitations can be viewed as the analogue in the Ising
case of the Berry phase effect in the spin-symmetric problem. This raises the question of the proper path integral of the spin-boson model
and related Ising-anisotropic models in the continuum limit.\\
In what follows, I point out the role of the Berry phase in the path integral representation of the spin-boson model.
%In what follows, I will discuss some of the subtleties underlying various  functional integral representations 
%of the spin-boson model and
%related models. 
%For simplicity, I focus on the spin-boson problem.
%of the two models in comparison 
%to the local $\phi^4$-theory\cite{}.
%\cite{Radcliffe.71,Inomata,Perelomov,Fradkin,Fahri.92,Kirchner.09,Zhu.04,Glossop.05,Fisher.72,Kirchner.08b,Rieger.99,Kirchner.09b}
%\section{The spin-boson model and path integral representations for spin} 

\begin{figure}
\begin{center} 
\includegraphics[%
  width=1.0\linewidth,
  keepaspectratio]{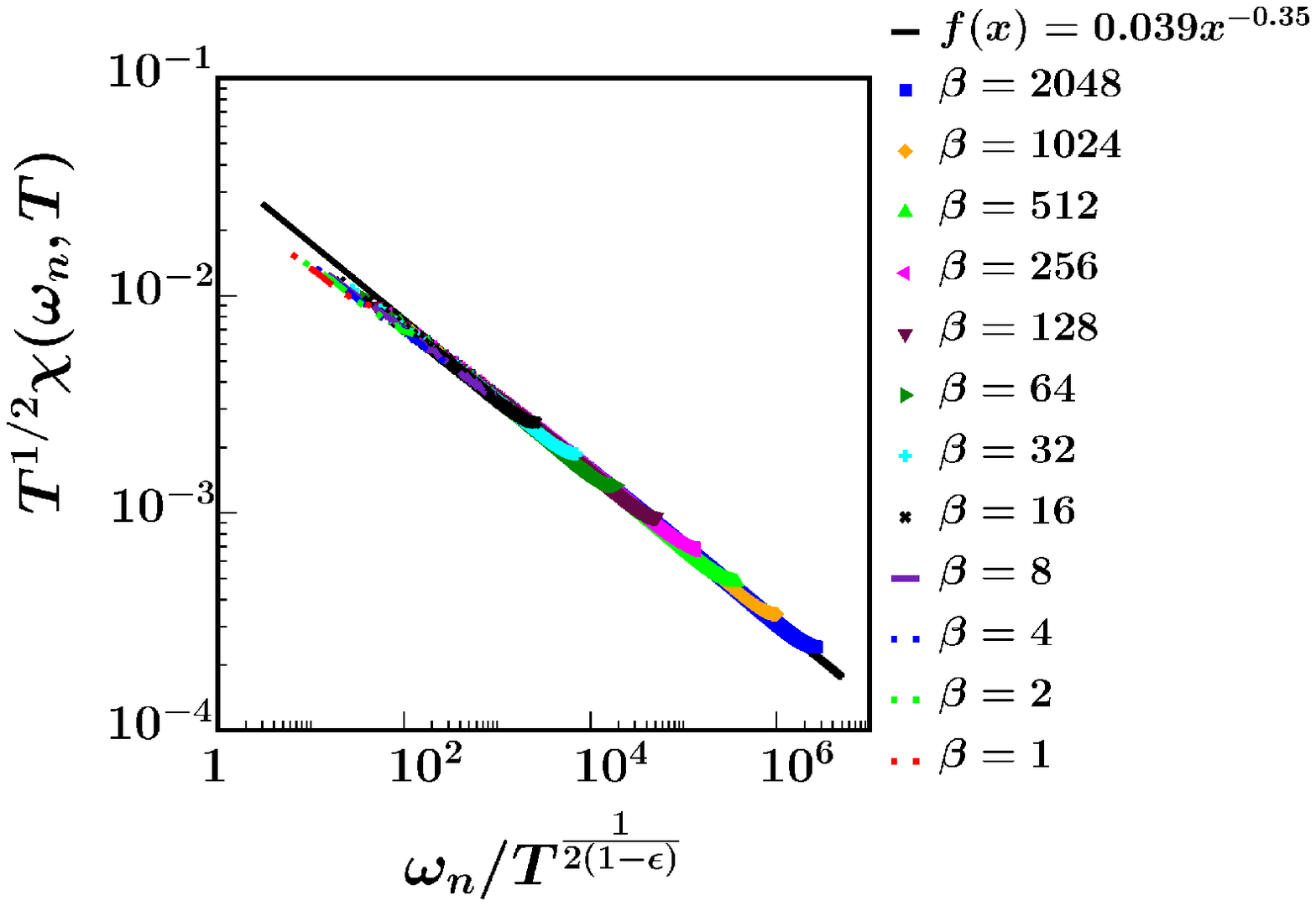}
\end{center}
\caption{Scaling function of the model corresponding to Eq.~(\ref{GLW}) for $\epsilon=0.65$ and $r_0=0$.
This value of $\epsilon$ places the model above its upper critical
dimension, resulting in $\chi_c(T=0,\omega_n)\sim |\omega_n|^{\epsilon-1}$ and $\chi_c(T,\omega_n=0)\sim T^{-1/2}$. 
Here, $\beta=1/T$ and $\omega_n=2\pi n/\beta$. Details of the algorithm that lead to the scaling plot can be found in
reference~\cite{Kirchner.09}.}
\label{scaling}
\end{figure}

The sub-Ohmic spin-boson model is defined by
\begin{equation}
  H=\Gamma\sigma^{x}+g\sigma^{z}\sum_{q}(a_q^{\dagger}+a^{}_q)+\sum_q \omega_q^{} a_q^{\dagger}a^{}_q,
\label{spinboson}
\end{equation}
where $\sigma^x$ and $\sigma^z$ are generators of the su(2) algebra and $a_q^{}$ and $a^{\dagger}_q$ follow bosonic
commutation relations. For a  sub-Ohmic spectral density of the bosons,
\begin{equation}
\label{spectraldensity}
\sum_q \big[\delta(\omega_{}^{}-\omega_q^{})-\delta(\omega_{}^{}+\omega_q^{})\big ]\sim |\omega|^{1-\eta}
\mbox{sgn}(\omega),~ \mbox{for}~~|\omega|<\Lambda,
\end{equation}
with $0<\eta<1$, the system
undergoes a continuous quantum phase transition at a critical coupling $g_c$ that separates a localized from a 
delocalized phase. It is well established that the associated critical exponents obey hyperscaling for $\eta<1/2$.
The situation for $1/2<\eta<1$ is less clear and has been discussed 
controversially~\cite{Glossop.07b,Kirchner.09,Vojta.05,Winter.09,Vojta.09}. 
This is so, because no reliable numerical method exists that allows to address
Eq.~(\ref{spinboson}) directly, despite the apparent simplicity of the spin-1/2 spin-boson model. 
The NRG
method e.g. relies on a truncation of the bosonic Hilbert space. 
The starting point of  a recent Monte Carlo study is an effective action description
for the spin degrees of freedom.  This
relies on a particular functional integral representation of the spin-boson partition function and the applicability of 
an integral transformation
\begin{equation}
\frac{\mbox{Det}[A]}{2\pi i}\int d\phi_i^* d\phi_j e^{-\phi_i^{*}A^{}_{i,j}\phi_j^{}+z_i^{}\phi_i^{*}+z_i^{*}\phi_i^{}}
=e^{z_i^{*}A^{-1}_{i,j}z_j^{}}
\label{integraltransform}
\end{equation}
for Gaussian integrals within the path integral representation in order to integrate out the bosons~\cite{Winter.09}.\\ 
A central question is, whether the effective action that results from integrating out the bosonic and fermionic degrees of freedom does assume
the form of an order parameter or Ginzburg-Landau-Wilson (GLW) functional. Such a GLW functional then serves as the 
starting point for a proper RG analysis. 
For the spin-boson problem, the expectation value of the
local spin operator $\sigma^z$ can be used as an order parameter. 
Ignoring any subtleties that arise when integrating out the bosons,
one may conclude that the effective action is of the form
\begin{equation}
\label{GLW}
S^{trial}_{eff}=\sum_{n} (r_0+\alpha |\omega_n|^{1-\eta}+\beta |\omega_n|^2)\phi_n\phi_{-n}+u\sum_{k,l,m,n}\phi_k\phi_l\phi_m\phi_n \delta_{k+l+m+n,0},
\end{equation} 
where $\phi_n$ is the Fourier component of the order parameter at the $n$th Matsubara frequency $\omega_n$.
%where $\omega_n$ are Matsubara frequencies, the $\phi_n$ is the Fourier components of the order parameter on the nth Matsubara frequency.
The term $|\omega_n|^{1-\eta}$ arises from a retarded long-ranged interaction $\sim |\tau|^{\eta-2}$, while   
$r_0$ measures the distance from the critical point and the term $\beta |\omega_n|^2$
represents a short-range interaction.
Assuming the validity of the quantum-to-classical mapping, this last term  is identified with $\Gamma \sigma^x$.
The action Eq.~(\ref{GLW}) is equivalent to a one-dimensional spin-chain with long-range interaction $\sim r^{\eta-2}$ and its critical properties 
are well studied~\cite{Fisher.72,Suzuki.76}. In particular, consider a RG rescaling that transforms $\omega_n'=b \omega_n$. 
Keeping the $\alpha$ term fixed implies
$\phi_{n}'=b^{(\eta-1)/2}\phi_n$. Therefore, $\beta$ changes as $b^{-1-\eta}$ and $u$ scales as $b^{1-2\eta}$. This implies that for
 $\eta>1/2$, 
$u$ scales to zero and the RG flow is governed by the Gaussian fixed point.
As a result, hyperscaling breaks down for $\eta>1/2$. This is  demonstrated in Figure \ref{scaling} for $\eta=0.65$,
where one finds  $\chi_c(T=0,\omega_n)\sim |\omega_n|^{\eta-1}$ and $\chi_c(T,\omega_n=0)\sim T^{-1/2}$ following reference~\cite{Kirchner.09}.
The fact, that $\eta-1\neq -1/2$ is a manifestation of the breakdown of hyperscaling.

In the following, some of the difficulties associated with the derivation of an effective local action are discussed and the need for spin coherent 
states in order to obtain a well-defined continuum limit is pointed out. A basic
introduction into spin coherent states is provided. The central result of this article is that the effective action of the spin-boson model does
not assume a form similar to Eq.~(\ref{GLW}) due to the presence of  a corresponding Berry phase term. 
The considerations above, based on Eq.~(\ref{GLW}), therefore do not readily apply to
the spin-boson model. In fact, a Berry phase term has been shown to change the critical behavior for $\eta>1/2$ in closely related 
models~\cite{Kirchner.08}. 
%question has become whether the effective action assumes the form of a local $\phi^4$-theory with $\epsilon=1/2$ 
%as its upper critical dimension, since this would imply that hyperscaling relations do not hold in general for  $1/2<\epsilon<1$.
%%%%%%%%%%%%%%%%%%% Figure 1 %%%%%%%%%%%%%%%%%%%%%%%%%%%%%%%%%%%%%%%%%%%%%%%%%%%%%%%%%%%%%%%%%
%%%%%%%%%%%%%%%%%%%%%%%%%%%%%%%%%%%%%%%%%%%%%%%%%%%%%%%%%%%%%%%%%%%%%%%%%%%%%%%%%%%%%%%%%%%%%%

\section{Path Integral Representations for Spin}
Path integrals have become a versatile tool in all areas of physics soon after
their introduction into quantum mechanics by Feynman in 1948. In semiclassical calculations e.g. one can bring
to bear its relation with the action of classical mechanics.  For a problem involving spin this relation
is less obvious and functional integral methods for spin systems are far less popular.
Nonetheless, a number of distinct functional integral representations for spin Hamiltonians can be found in the 
literature~\cite{Radcliffe.71,Fahri.92,Inomata,Nielson.88,Stone.89,Rieger.99,Kuratsuji.80}. 
For the spin-boson problem and in particular for constructing its effective action
in the reduced Hilbert space of  spin states not all functional integral representations 
are equally well suited.
Fahri and Gutmann e.g. derived a functional integral directly from the Hamiltonian based on an orthogonal, 
countable basis of the Hilbert space on which the Hamiltonian operates~\cite{Fahri.92}.
The spectrum of Eq.~(\ref{spinboson}) is continuous so that the Fahri-Gutmann representation does not apply to the sub-Ohmic
spin-boson problem. Furthermore, since paths are constructed  as continuous Markov chains in the orthogonal basis, 
the resulting measure is in general non-Gaussian for the bosonic modes and the integral transformation, 
Eq.(\ref{integraltransform}) is therefore not readily applicable. It therefore is far from obvious
if or how an  effective action could be obtained from the Fahri-Gutmann 
functional integral. 
To ensure the applicability of Eq.~(\ref{integraltransform}), the bosonic operators appearing in the functional integral of the 
spin-boson model, Eq.~(\ref{spinboson}),  will be
represented by bosonic coherent states that diagonalize the bosonic annihilation operator 
$a|\phi\rangle=\phi|\phi\rangle$; where $\phi$ is a complex number, $\phi \, \in \,\mathbb{C}$.  
It can be shown that the overlap between bosonic coherent states is
\begin{equation}
\langle\tilde{\phi}|\tilde{\phi}^{'}\rangle=e^{-|\tilde{\phi}-\tilde{\phi}^{'}|^2}\leq 1,
\end{equation} 
for properly normalized coherent states such that $\langle \tilde{\phi}|\tilde{\phi}\rangle=1$.
The resolution of unity in the single mode Hilbert space ${\mathcal{H}}_B$ in terms of the $|\phi \rangle$ becomes
\begin{equation}
\mathbf{1_{\phi}}=\frac{1}{2\pi i}\int d\phi^{*}d\phi\: e^{-\phi^{*}\phi}|\phi\rangle \langle\phi|.
\end{equation} 
For brevity, we will in this section only consider the Hamiltonian
\begin{equation}
H_{SB}=\Gamma\sigma^{x}+g\sigma^{z}(a^{\dagger}+a)+\omega a^{\dagger}a,
\label{simplified}
\end{equation}
since the extension to the full sub-Ohmic spin-boson problem will not pose any additional difficulties.
The Hilbert space of the spin part of Eq.~(\ref{simplified}), ${\mathcal{H}}_S$,  is spanned by the eigenstates of $\sigma^{z}$:
$|S,m\rangle,~m \varepsilon \{-S,-S+1,\ldots, S\}$.
For a spin-1/2 system, $S=1/2$, the resolution of unity in  ${\mathcal{H}}_S$ becomes
\begin{equation}
\mathbf{1}_{\sigma}=\sum_{\sigma=\uparrow,\downarrow}|\sigma\rangle \langle\sigma|=|\uparrow\rangle \langle \uparrow|+|\downarrow \rangle \langle\downarrow|,
\end{equation}
and the matrix elements of $\sigma^x$ follow from $\sigma^{x}|\uparrow\rangle=|\downarrow\rangle,~\sigma^{x}|\downarrow\rangle=|\uparrow \rangle$.
A basis of the Hilbert space of  $H_{SB}$, ${\mathcal{H}}_B\otimes {\mathcal{H}}_S$, is therefore given by $|\phi\sigma\rangle=|\phi\rangle|\sigma\rangle$, where
$\phi$ is an arbitrary complex number and $\sigma$ is either $\uparrow$ or
$\downarrow$. The partition function $\mathcal{Z}$ of $H_{SB}$ therefore can be written as 
\begin{eqnarray}
\label{partition1}
\mathcal{Z} & = & \mbox{Tr}\left\{ e^{-\beta H}\right\} 
 = \frac{1}{2\pi i}\sum_{\sigma=\uparrow,\downarrow}\int d\phi^{*}d\phi\, e^{-\phi^{*}\phi}\langle\sigma\phi|e^{-\beta H}|\sigma\phi \rangle 
\nonumber \\
 & = &A\prod_{i}^{M}\sum_{\sigma_{i}=\uparrow,\downarrow}\int d\phi_{i}^{*}d\phi_{i}^{}e^{-\phi_{i}^{*}\phi_{i}^{}}\langle\sigma_{M}\phi_{M}|e^{-\epsilon H}|\sigma_{1}\phi_{1}\rangle\\
 & \times& \langle\sigma_{1}\phi_{1}|\ldots|\sigma_{i}\phi_{i}\rangle \langle\sigma_{i}\phi_{i}|e^{-\epsilon H}|\sigma_{i+1}\phi_{i+1}\rangle\nonumber \\
 & \times& \langle\sigma_{i+1}\phi_{i+1}|\ldots|\sigma_{M-1}\phi_{M-1}\rangle \langle\sigma_{M-1}\phi_{M-1}|e^{-\epsilon H}|\sigma_{M}\phi_{M}\rangle,\nonumber
\end{eqnarray}
where $A$ is a  constant and $\epsilon=\beta/M$. 
Since $\epsilon \ll 1$, each of the matrix elements $\langle\sigma_{i}\phi_{i}|e^{-\epsilon H}|\sigma_{i+1}\phi_{i+1}\rangle$ appearing in Eq.~(\ref{partition1}) 
can be further simplified:
\begin{eqnarray}
\label{shorttime1}
\langle\sigma_{i}\phi_{i}|e^{-\epsilon H}|\sigma_{i+1}\phi_{i+1}\rangle & = & \langle \sigma_{i}\phi_{i}|1-\epsilon H+O(\epsilon^{2})|\sigma_{i+1}\phi_{i+1}\rangle \\
% & = & <\sigma_{i}\phi_{i}|1-\epsilon[\Gamma\sigma^{x}+\omega a^{\dagger}a+g\sigma^{z}(a^{\dagger}+a)]|\sigma_{i+1}\phi_{i+1}>\\
 & = & \langle\phi_{i}|\phi_{i+1}\rangle \Big(\langle\sigma_{i}|\sigma_{i+1}\rangle-\epsilon[\Gamma\langle\sigma_{i}|\sigma^{x}|\sigma_{i+1}\rangle\nonumber \\
&+&\omega\phi_{i}^{*}\phi_{i+1}\langle\sigma_{i}|\sigma_{i+1}\rangle
 +g\langle\sigma_{i}|\sigma^{z}|\sigma_{i+1}\rangle(\phi_{i}^{*}+\phi_{i+1})]\Big).\nonumber 
\end{eqnarray}
The spin basis is orthonormal, $\langle\sigma_{i}|\sigma_{i+1}\rangle=\delta_{\sigma_{i},\sigma_{i+1}}$. 
Therefore,
the matrix element can be rewritten as
\begin{eqnarray}
\label{matrixelement1}
\langle\sigma_{i}\phi_{i}|e^{-\epsilon H}|\sigma_{i+1}\phi_{i+1}\rangle 
%& = & <\phi_{i}|\phi_{i+1}>(<\sigma_{i}|\sigma_{i+1}>-\epsilon[\Gamma<\sigma_{i}|\sigma^{x}|\sigma_{i+1}>+\omega\phi_{i}^{*}\phi_{i+1}\delta_{i,i+1}\\
% &  & +g\sigma_{i+1}^{z}<\sigma_{i}|\sigma_{i+1}>(\phi_{i}^{*}+\phi_{i+1})])\\
% & = & <\phi_{i}|\phi_{i+1}>(\delta_{i,i+1}-\epsilon[\Gamma<\sigma_{i}|\sigma^{x}|\sigma_{i+1}>+\omega\phi_{i}^{*}\phi_{i+1}\delta_{i,i+1}\\
% &  & +g\sigma_{i+1}^{z}\delta_{i,i+1}(\phi_{i}^{*}+\phi_{i+1})])\\
 & = & \langle\phi_{i}|\phi_{i+1}\rangle \Big( \delta_{\sigma_{i},\sigma_{i+1}}-\epsilon[\Gamma\langle\sigma_{i}|\overline{\sigma}_{i+1}\rangle\nonumber\\
& & +\omega\phi_{i}^{*}\phi_{i+1}\delta_{\sigma_{i},\sigma_{i+1}}
+g\sigma_{i+1}^{z}\delta_{\sigma_{i},\sigma_{i+1}}(\phi_{i}^{*}+\phi_{i+1})]\Big )\nonumber\\
 & = & \langle\phi_{i}|\phi_{i+1}\rangle \Big( \delta_{\sigma_{i},\sigma_{i+1}}-\epsilon[\Gamma\delta_{\sigma_{i},\overline{\sigma}_{i+1}}\\
& & +\omega\phi_{i}^{*}\phi_{i+1}\delta_{\sigma_{i},\sigma_{i+1}}
+g\sigma_{i+1}^{z}\delta_{\sigma_{i},\sigma_{i+1}}(\phi_{i}^{*}+\phi_{i+1})]\Big ),\nonumber
\end{eqnarray}
 where $|\overline{\sigma}_{i+1}\rangle=|\uparrow\rangle$ if $|\sigma_{i+1}\rangle=|\downarrow\rangle$
or $|\overline{\sigma}_{i+1}\rangle=|\downarrow\rangle$ if $|\sigma_{i+1}\rangle=|\uparrow\rangle$ and $\sigma^{z}_{i+1}=\pm 1$ depending on the spin state $|\sigma_{i+1}\rangle$.
In the final form of the matrix element, Eq.~(\ref{matrixelement1}), no operators appear on the right hand side. As a consequence of the orthonormal
spin basis, Eq.~(\ref{matrixelement1}) assumes the form of a matrix exponential:
\begin{eqnarray}
\label{matrixP}
\langle\sigma_{i}\phi_{i}|e^{-\epsilon H}|\sigma_{i+1}\phi_{i+1}\rangle &=& e^{\phi_{i}^{*}\phi_{i+1}} \times \\
%<\phi_{i}|\phi_{i+1}>(\delta_{i,i+1}-\epsilon[\Gamma\delta_{i,\bar{i+1}}+\omega\phi_{i}^{*}\phi_{i+1}\delta_{i,i+1}+g\sigma_{i+1}^{z}\delta_{i,i+1}(\phi_{i}^{*}+\phi_{i+1})])\\
 &  & \hspace*{-3.0cm}\left(\begin{array}{cc}
1-\epsilon[\omega\phi_{i}^{*}\phi_{i+1}+g(\phi_{i}^{*}+\phi_{i+1})] & -\epsilon\Gamma \nonumber\\
-\epsilon\Gamma & 1-\epsilon[\omega\phi_{i}^{*}\phi_{i+1}-g(\phi_{i}^{*}+\phi_{i+1})]\end{array}\right)\nonumber \\
% & = & e^{\phi_{i}^{*}\phi_{i+1}}\left(\left(\begin{array}{cc}
%1 & 0\\
%0 & 1\end{array}\right)-\epsilon\left(\begin{array}{cc}
%\omega\phi_{i}^{*}\phi_{i+1}+g(\phi_{i}^{*}+\phi_{i+1}) & \Gamma\\
%\Gamma & \omega\phi_{i}^{*}\phi_{i+1}-g(\phi_{i}^{*}+\phi_{i+1})\end{array}\right)\right)\\
% & = & e^{\phi_{i}^{*}\phi_{i+1}}\left(1-\epsilon\mathbf{M}\right)\\
& \equiv & e^{\phi_{i}^{*}\phi_{i+1}} {\bm P}^{i,i+1}=e^{\phi_{i}^{*}\phi_{i+1}}e^{-\epsilon \mathbf{K}^{i,i+1}}, \nonumber
\end{eqnarray}
where matrix ${\bm P}^{i,i+1}=\exp (-\epsilon \mathbf{K}^{i,i+1})$ is defined for later use and ${\bm K}^{i,i+1}$ is the matrix 
\begin{eqnarray}
\label{expo}
\mathbf{K}^{i,i+1}&=&\left(\begin{array}{cc}
\omega\phi_{i}^{*}\phi_{i+1}+g(\phi_{i}^{*}+\phi_{i+1}) & \Gamma \\
\Gamma & \omega\phi_{i}^{*}\phi_{i+1}-g(\phi_{i}^{*}+\phi_{i+1})\end{array}\right) \\
&=& \omega\phi_{i}^{*}\phi_{i+1}{\bm 1}  +\Gamma {\bm \sigma}^{x} +g  (\phi_{i}^{*}+\phi_{i+1}){\bm \sigma}^{z}.
\label{expo2}
\end{eqnarray}
Note, the Pauli matrices appearing in Eq.~(\ref{expo}) do not carry a time label and  the $i$- and $(i+1)$-dependence of $\mathbf{K}^{i,i+1}$ is entirely due to
the imaginary time dependence of the $\phi$ variable characterizing the boson state
 that enters the prefactors of $\bm 1$ and ${\bm \sigma}^z$ of Eq.~(\ref{expo2}). 
The change in spin state during the time interval from $i$ to $i+1$
is determined by choosing one of the four matrix elements in Eq.~(\ref{expo}). The structure of the matrix is the same for each of the $M$ time intervals 
but the diagonal elements depend on the time label $i$ through the boson.
Inserting  Eq.~(\ref{expo}) into  Eq.~(\ref{partition1}), the continuum limit $\epsilon \rightarrow 0$ cannot easily be performed, since
one deals with  matrix exponentials of non-commuting matrices. 
As a consequence, this representation does not allow to globally 
integrate out the bosonic degrees of freedom in a straightforward manner.
Simply ignoring the complications that arise from the non-commuting matrices and taking $\epsilon \rightarrow 0$ and applying Eq.~(\ref{integraltransform})
%Applying the integral transformation, Eq.~(\ref{integraltransform}), instead directly to Eq.~(\ref{matrixP}), 
would lead to terms of the 
form ${\bm \sigma}^z{\bm \sigma}^z=\bm 1$, since the Pauli matrices do not carry a time label. 
If the coupling to the bosonic degrees of freedom vanishes, $g=0$, the matrix $\bm K$ at different times
commutes and the continuum limit can be performed.
\newline
An apparent alternative seems to be to rewrite the matrix elements of $\bm P$ in  Eq.~(\ref{matrixP}) as the exponential of some function $F(S_i,S_{i+1})$
with $S_i=\pm 1$:
 \begin{equation}
\label{versuch2}
\bm P|_{S_i,S_{i+1}}=e^{a+b(S_i+S_{i+1})+c S_i S_{i+1}}.
\end{equation}
This leads to
\begin{eqnarray*}
a&=&\frac{1}{2}\Big(-\epsilon \omega \phi^{*}_{i+1}\phi^{}_i+\ln(-\epsilon \Gamma)\Big),\\
b&=&-\frac{1}{2}\epsilon g \Big (\phi^{*}_{i+1}+\phi^{}_i \Big),\\
c&=&\frac{1}{2}\Big(-\epsilon \omega \phi^{*}_{i+1}\phi^{}_i-\ln(-\epsilon \Gamma)\Big).
\end{eqnarray*}
Eq.~(\ref{versuch2}) can be inserted in Eq.~(\ref{partition1}) for the partition function. As is clear from e.g. the terms coupling to
$ S_i S_{i+1}$, the continuum limit $\epsilon \rightarrow 0$ can again not easily be taken.
In particular, the bosonic field couples to the nearest neighbor coupling term $S_i S_{i+1}$ through $c$ of Eq.~(\ref{versuch2}),
that otherwise would give rise to a short-range coupling as in Eq.~(\ref{GLW}). 
%Simply integrating out the bosonic field in the
%short time propagator is not possible either, since $c$ in Eq.(\ref{versuch2}) depends on $ \phi^{*}_{i+1}\phi^{}_i$ resulting in a quartic
%coupling of $S$ and $\phi$.
The difficulties encountered in the path integral representation above are related to the orthogonality of the spin basis in 
${\mathcal{H}}_S$ while the basis in ${\mathcal{H}}_B$ has the property $\langle\phi|\phi'\rangle \neq 0$. 
If the coupling to the bosons vanishes ($g=0$), Eq.~(\ref{versuch2}) reduces to the standard transfer-matrix of a quantum 
spin in a transverse field $\Gamma$ and standard methods apply~\cite{Krzakala.08}.\\ 
%The possibility and the resulting difficulties,
%to choose an orthonormal basis also in ${\mathcal{H}}_B$ following Fahri and Gutmann~\cite{Fahri.92}
%have briefly been discussed above. 
As noted above, a strategy of choosing orthonormal bases for both ${\mathcal{H}}_S$ and ${\mathcal{H}}_B$ is problematic because Eq.~(\ref{integraltransform})
would no longer be applicable.
A more viable option lies in the
construction of an overcomplete basis in ${\mathcal{H}}_S$ in close analogy to the bosonic coherent states. 
These states are therefore called spin coherent states.

\section{Spin Coherent States}

Spin coherent states were first introduced by J.~Radcliffe in 1971~\cite{Radcliffe.71} and are  discussed e.g. in 
\cite{Inomata,Perelomov,Kuratsuji.80,Fradkin}.
Let us start considering the spin-1/2 problem. The physical state of such a two-level system is specified by a state vector
\begin{equation}
  |\Psi\rangle=\alpha |\uparrow\rangle +\beta|\downarrow\rangle,~\mbox{with}~|\alpha|^2+|\beta|^2=1,
\end{equation}
up to a phase; the Hilbert space is spanned by the two basis states $|\downarrow\rangle,~ |\uparrow\rangle$ with $\langle\uparrow|\downarrow\rangle=0$.
Thus, the one-dimensional projector
\begin{eqnarray}
\label{projector}
{\mathcal{P}}_{\Psi}&=& |\Psi\rangle\langle\Psi|=\left(\begin{array}{cc}
|\alpha|^{2} & \alpha \beta^{*} \\
\alpha^{*} \beta & |\beta|^{2}\end{array}\right)
=\frac{n_x}{2} {\bm \sigma}^x+\frac{n_y}{2} {\bm \sigma}^y +\frac{n_z}{2} {\bm \sigma}^z+ t {\bm 1},
\end{eqnarray}
with $\mbox{Tr}({\mathcal P}_{\Psi})=1$ 
%and $\mbox{Det}({\mathcal P}_{\Psi})=0$ 
completely specifies the state of the two-level system. Here, ${\bm \sigma}^x,~{\bm \sigma}^y,~{\bm \sigma}^z$ are the Pauli
matrices and ${\bm 1}$ is the $2\times 2$ unit matrix.
It follows that $n_x^2+n_y^2+n_z^2=1$ and ${\mathcal P}_{\Psi}$ is mapped onto a  point $P=(n_x,n_y,n_z)$ on a sphere of 
radius $1$, 
while the orthogonal state vector 
${\mathcal Q}_{\Psi}=1-{\mathcal P}_{\Psi}$ corresponds to the point $Q=(-n_x,-n_y,-n_z)=-P$ of the Bloch sphere.  
Any point 
$P'=(n_x',n_y',n_z')$ on the sphere can be mapped onto any other point  $P=(n_x,n_y,n_z)$ via a similarity transform,
\begin{eqnarray}
{\mathcal{P}}_{\Psi}= {\bm V}{\mathcal{P}}_{\Psi^{'}} {\bm V}^{-1},
\end{eqnarray}
where ${\bm V}$ is a member of the group SU(2). 
%and ${\bm V}^{-1}$ are SU(2) group members.
The dynamics of the two-level system are described by paths on the Bloch sphere generated by the time evolution operator associated with
a particular Hamiltonian. The spin part of such a Hamiltonian is an element of the su(2) algebra and generates a mapping 
into SU(2) via the exponential map. 
Apparently, $ {\bm V}={\bm 1}$ and  ${\bm V}=-{\bm 1}$ leave  ${\mathcal{P}}_{\Psi}$ invariant. 
This is a manifestation of the adjoint representation
of SU(2) on which the mapping onto the Bloch sphere builds.

\begin{figure}
\begin{center} 
\includegraphics[%
  width=1.0\linewidth,
  keepaspectratio]{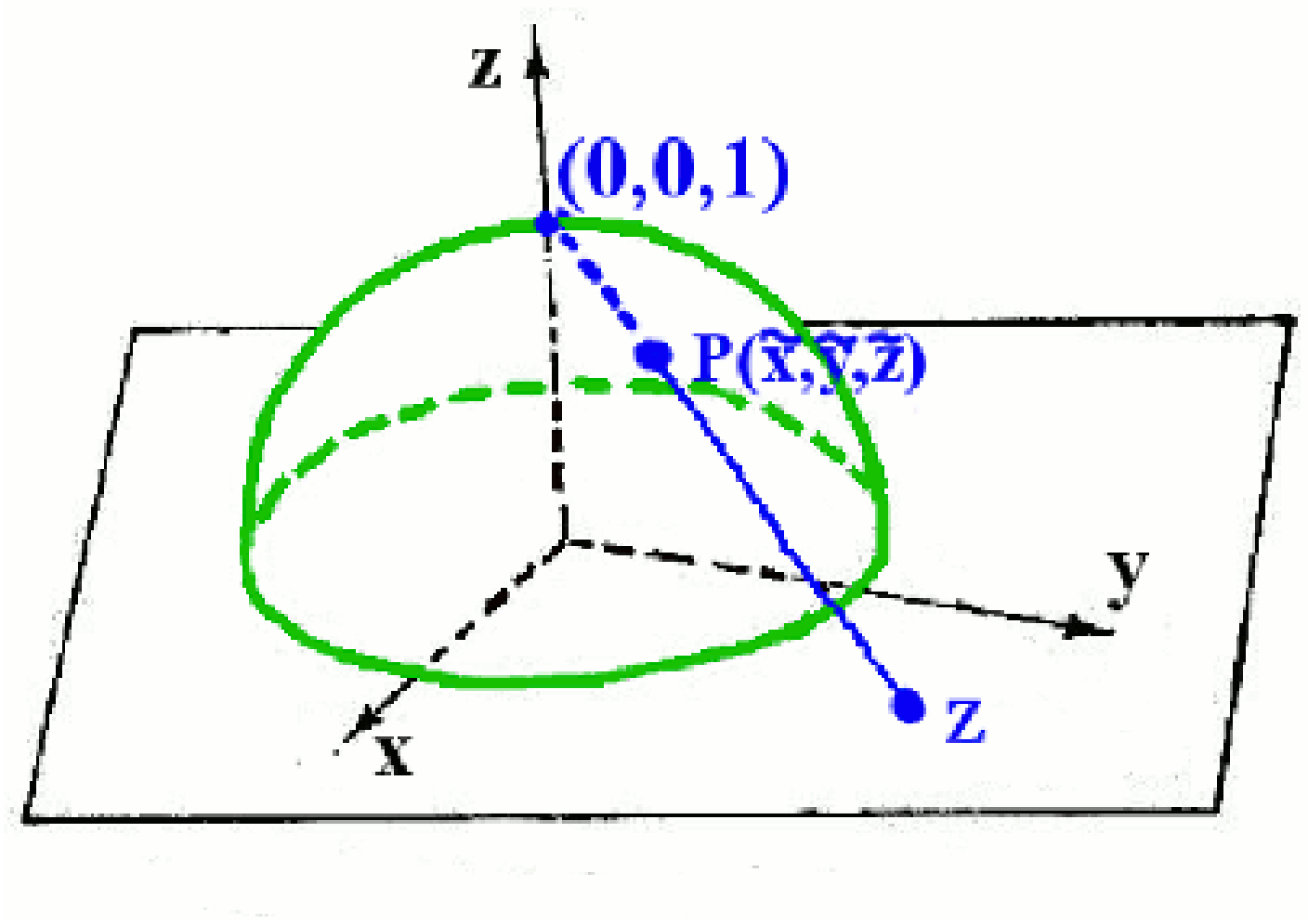}
\end{center}
\caption{Stereographic projection of the unit sphere to the complex plane: The coordinates of the point 
$P=(\tilde{x},\tilde{y},\tilde{z})$
are mapped onto the point $Z=(x,y)=(\tilde{x}/(1-\tilde{z}),\tilde{y}/(1-\tilde{z}))$ in the plane. 
The coordinates of $P$ in terms of $x,y$ are 
$\Big ( \tilde{x},\tilde{y},\tilde{z}\Big )=\Big(2x/(1+x^2+y^2),2y/(1+x^2+y^2),(x^2+y^2-1)/(1+x^2+y^2\Big)$.}
\label{compare}
\end{figure}

In order to define spin-coherent states, it is useful to note that 
if the $|\downarrow\rangle$-state, corresponding to the south pole of the sphere, is chosen as a reference state, only the north pole  has a
vanishing overlap with the reference state.  Any other point on the sphere
will correspond to a state that has a non-vanishing overlap with $|\downarrow\rangle$. This suggests, 
to define a spin-coherent state $|z\rangle$ via\cite{Radcliffe.71,Inomata}.
\begin{equation}
\label{spincoherent}
|z\rangle=c e^{z J^{+}}|\downarrow\rangle=c(|\downarrow\rangle+z |\uparrow\rangle),
\end{equation}
in close analogy to the definition of bosonic coherent states.
%In close analogy to bosonic coherent states we therefore define the spin coherent
%states for spin-1/2 as
where $z\in \mathbb{C}$ is a complex number and $c=(1+|z|^2)^{-1/2}$ follows from $\langle z|z\rangle=1$ and $J^{+}|\uparrow\rangle=0$.
Consequently, the state $|z\rangle=|x+iy\rangle$ is represented by a point $(x,y)$ in the complex plane as well as 
by its image  $P=(n_x,n_y,n_z)$ on the Bloch sphere (also called $S^2$). 
The relation between $z$ in the complex plane and $P$ on $S^2$ follows from Eq.~(\ref{projector}) and  Eq.~(\ref{spincoherent})
and is just the stereographic projection, see Figure \ref{compare}:
\begin{eqnarray}
n_x=\frac{2 x}{1+|z|^2},~n_y=\frac{2 y}{1+|z|^2},~n_z=\frac{|z|^2-1}{1+|z|^2}.
\end{eqnarray}
In the remainder of this section, the formulae necessary to set up the spin path integral are valid for a general representation $S$
of SU(2).

Recall, that for any irreducible representation $|S,m\rangle$ of SU(2),
\begin{eqnarray}
\label{SM1}
J^{\pm}|S,m \rangle&=& \sqrt{(S\mp m)(S\pm m+1)}|S, m\pm 1 \rangle\\
J^z|S,m \rangle&=& m |S,m \rangle \\
\sum_{m=-S}^{S} |S,m\rangle \langle S,m|&=& 1.
\label{SM2} 
\end{eqnarray}  
In accordance with Eq.~(\ref{spincoherent}), the definition of the spin coherent state for general $S$ is
\begin{eqnarray}
|z\rangle =c e^{z J^{+}}|S,-S \rangle&=&c(\sum_{l=0}^{2S}\frac{z^l}{l!}(J^{+})^l|S,-S \rangle)\nonumber \\
&=&c \sum_{l=0}^{2S} z^l \sqrt{\frac{(2S)!}{l!(2S-l)!}}|S,-S+l \rangle,
\label{spincoherentallg}
\end{eqnarray} 
where  Eq.(\ref{SM1}) was used. From the orthogonality of the spin states, 
the normalization factor $c$ follows as $c=(1+|z|^2)^{-S}$ and the expansion coefficient of $|z\rangle$ in terms of the 
$|S,m \rangle$ becomes
\begin{equation}
\langle S,m|z \rangle=\left( {\begin{array}{*{20}c} 2S \\ S+m \\ \end{array}} \right)\frac{z^{S+m}}{(1+|z|^2)^{S}}.
\end{equation} 
%%%%%%%%%%%%%%%%%%%%%%%%%%%%%%%%%%%%%%%%%% Figure 2 %%%%%%%%%%%%%%%%%%%%%%%%%%%%%%%%%%%%%%%%%%%%%%%%%%%%%%%%%%%
%%%%%%%%%%%%%%%%%%%%%%%%%%%%%%%%%%%%%%%%%%%%%%%%%%%%%%%%%%%%%%%%%%%%%%%%%%%%%%%%%%%%%%%%%%%%%%%%%%%%%%%%%%%%%%%%
The resolution of unity for spin coherent states is given by
\begin{equation}
\label{unity}
\mathbf{1}_{}=\int dzdz^{*}\frac{2S+1}{2\pi(1+|z|^{2})^{2}}|z\rangle \langle z|.
\end{equation}
The proof of this equation is analogous to the one for boson coherent states\cite{Negele}. 
Using Eq.~(\ref{spincoherentallg}) and with the help of
\begin{equation}
  \frac{1}{2\pi}\int dz^{*}dz\, \frac{(|z|^2)^n}{(1+|z|^2)^{2S+2}}=\frac{\Gamma(1+n)\Gamma(1-n+2S)}{\Gamma(2+2S)},
\end{equation}
Eq.~(\ref{unity}) is shown to be equivalent to Eq.~(\ref{SM2}).  
Alternatively, one can make use of Schur's first lemma according to which an operator must be a constant if this operator 
commutes with every element
of an irreducible representation of SU(2)~\cite{Negele}. Hence, this operator must be proportional to the unit 
operator.
For general $S>1/2$, the pure state space is not the Bloch sphere $S^2$. Nonetheless, since $S^2$ is the coset space of $SU(2)$,
it will turn out that the spin path integral can be either formulated in the complex plane, using Eq.~(\ref{unity}),
or on $S^2$. From Eq.~(\ref{unity}) and the stereographic projection, the measure
\begin{equation}
d\mu=dzdz^{*}\frac{2S+1}{2\pi(1+|z|^{2})^{2}}
\end{equation}
can be interpreted as the area element of the unit sphere $S^2$. This measure is the invariant measure
over SU(2).

As mentioned above, spin coherent states are overcomplete and the overlap between spin coherent states $|z_1 \rangle$ and
$|z_2 \rangle$ is
\begin{equation}
\label{overlap}
\langle z_{1}|z_{2} \rangle=\frac{(1+z_{1}^{*}z_{2})^{2S}}{(1+|z_{1}|^{2})^{S}(1+|z_{2}|^{2})^{S}},
\end{equation}
this follows directly from Eq.~(\ref{spincoherent}).
The matrix elements of the spin operators $J^x$, $J^y$, and $J^z$ in the spin coherent state representation can
be calculated from  Eq.~(\ref{spincoherent}):
\begin{eqnarray}
\label{MEJx}
\frac{\langle z_1|J^{+}|z_2 \rangle}{\langle z_1|z_2 \rangle}&=&S\frac{2 z_1^{*}}{1+ z_1^{*}z_2^{}},\\
\frac{\langle z_1|J^{-}|z_2 \rangle}{\langle z_1|z_2 \rangle}&=&S\frac{2 z_2^{}}{1+ z_1^{*}z_2^{}},\\
\frac{\langle z_1|J^{z}|z_2 \rangle}{\langle z_1|z_2 \rangle}&=&S\frac{z_1^{*}z_2^{}-1}{1+ z_1^{*}z_2^{}}.
\label{MEJz}
\end{eqnarray}
For the diagonal part, one therefore obtains
\begin{eqnarray}
\frac{\langle z|J^{x}|z \rangle}{\langle z|z\rangle}&=&S\frac{2 \mbox{Re}(z)}{1+ |z|^2}=S n_x,\\
\frac{\langle z|J^{y}|z \rangle}{\langle z|z\rangle}&=&S\frac{2 \mbox{Im}(z)}{1+ |z|^2}=S n_y,\\
\frac{\langle z|J^{z}|z \rangle}{\langle z|z\rangle}&=&S\frac{|z|^2-1}{1+ |z|^2}=S n_z,
\end{eqnarray}
where $\vec{n}=(n_x,n_y,n_z)$ is the point $P$ on the unit sphere $S^2$ onto which $z$ is mapped by the stereographic projection. 
The definition of the spin coherent state for general spin representation $S$ suggests, that a close relation 
might exist 
with bosonic coherent states for the case, where the series in Eq.(\ref{spincoherentallg}) does not truncate, e.g. for
$S\rightarrow \infty$. This limit of  large S for the quantum critical point of the sub-Ohmic Bose-Fermi Kondo and spin-Boson models 
has been  explored in \cite{Kirchner.08e,unpl}.

Note, we did not assume any particular symmetry property of a  spin Hamiltonian.
All the properties discussed in this section are consequences of the various basis states for the Hilbert space  
of a quantum spin and will be valid for any spin Hamiltonian, spin-isotropic or not. 
%all that is needed is that $H_{S}$ is a member of the su(2) algebra. Through the exponential map, it therefore will
%generate some unitary time 
%evolution that takes the system from one point of the Bloch sphere to another or equivalently from one point of the complex 
%plane to another. The north pole of the Bloch sphere then corresponds to complex infinity.   

\section{Spin Path Integral Representation of the Spin-Boson Model}

\begin{figure}
\begin{center} 
\includegraphics[%
  width=0.6\linewidth,
  keepaspectratio]{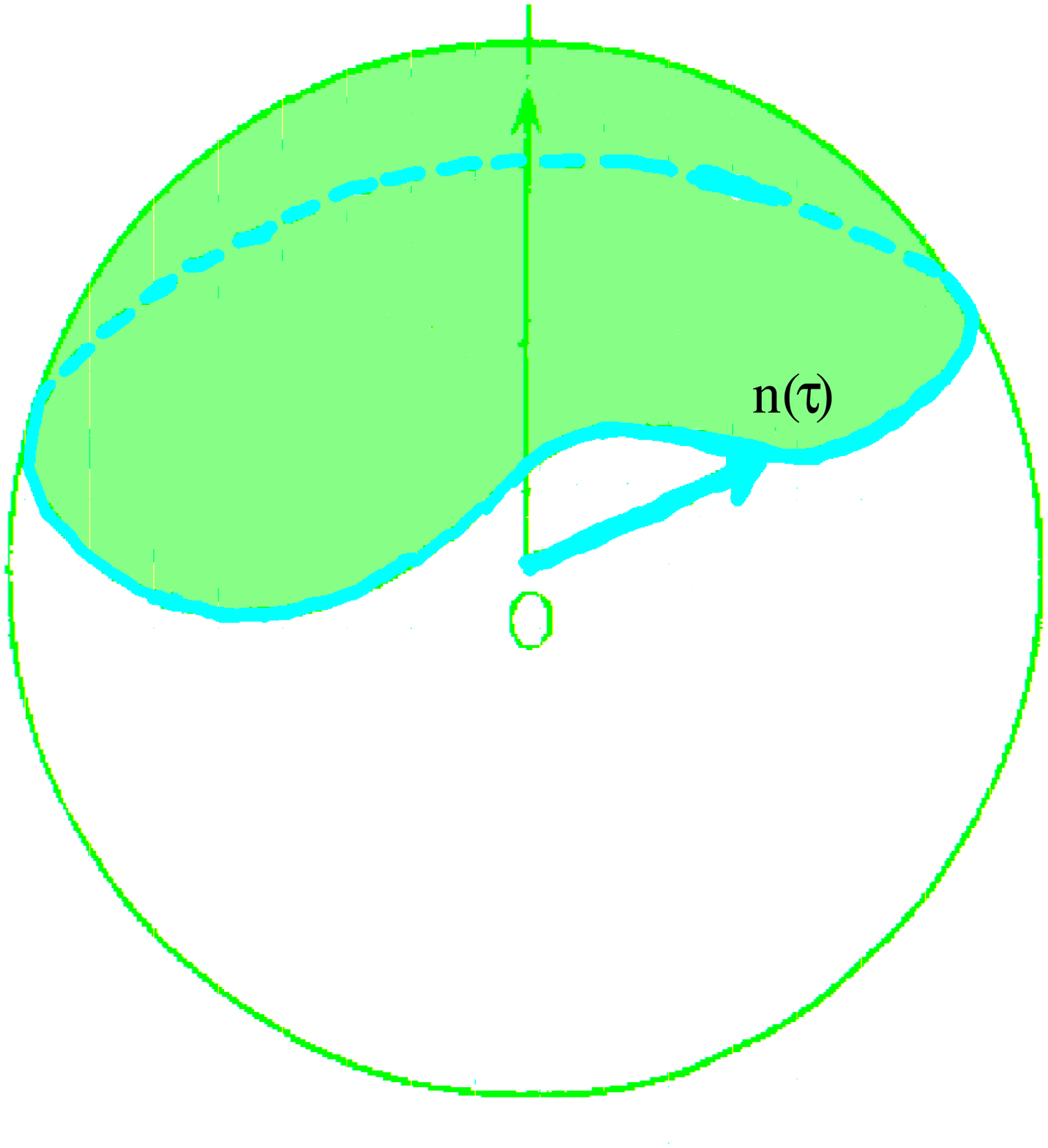}
\end{center}
\caption{The area ${\mathcal A}[\vec{n}]$ for  a particular path $\vec{n}(\tau)$:
${\mathcal A}[\vec{n}]$ is the shaded area traced out by the closed path $\vec{n}(\tau)$. 
The Berry phase associated with $\vec{n}(\tau)$
is given by $e^{i S {\mathcal A}[\vec{n}]}$.}
\label{berryphase}
\end{figure}

This section contains the  derivation of the functional integral representation of the sub-Ohmic
spin-boson model in terms of bosonic and spin coherent states and a short
discussion of the so-called Berry phase of the spin path integral.
We start again with 
\begin{equation}
\label{Hamiltonian}
H=\Gamma\sigma^{x}+g\sigma^{z}(a^{\dagger}+a)+\omega a^{\dagger}a.
\end{equation}
The resolution of unity in terms of boson and spin coherent states in the respective sub-spaces are
\begin{equation}
\mathbf{1_{\phi}}=\frac{1}{2\pi i}\int d\phi^{*}d\phi\: e^{-\phi^{*}\phi}|\phi\rangle \langle\phi|,
\end{equation}
and
\begin{equation}
\mathbf{1}_{\sigma}=\int dzdz^{*}\frac{2S+1}{2\pi(1+|z|^{2})^{2}}|z\rangle \langle z|
\end{equation}
respectively. 
A basis of the full Hilbert space is given by $|\phi z\rangle$ where $\phi$
and $z$ are complex numbers, $\phi,z \, \in \,\mathbb{C}$ .
%The spin coherent states obey
%
%\[
%<z_{i}|z_{i+1}>=\frac{(1+z_{i}^{*}z_{i+1})^{2S}}{(1+|z_{i}|^{2})^{S}(1+|z_{i+1}|^{2})^{S}}\]
%
%where $S$ labels the choosen irreducible representation (to be derived).
Repeating the steps above in the derivation of the partition function
in terms of a path integral in imaginary time yields:
\begin{eqnarray}
\label{partitionF}
\mathcal{Z} & = & \mathcal{T}r\left\{ e^{-\beta H}\right\} \\
% & = & \frac{1}{2\pi i}\:\int dzdz^{*}\int d\phi^{*}d\phi\, e^{-\phi^{*}\phi}\frac{2S+1}{2\pi(1+|z|^{2})^{2}}\langle z\phi|e^{-\beta H}|z\phi \rangle \nonumber \\
 & = & \frac{2S+1}{4\pi^2 i}\:\int dz_M^{}dz^{*}_M\int d\phi^{*}_Md\phi^{}_M\,
\frac{e^{-\phi^{*}_M\phi^{}_M}}{(1+|z_M|^{2})^{2}}\langle z_M\phi_M|(e^{(-\beta/M)H})^{M}|z_M\phi_M \rangle.\nonumber
\end{eqnarray}
Inserting $(M-1)$ times the resolution of unity ${\bm 1}_\phi \otimes {\bm 1}_{\sigma}$ in the product space yields 
\begin{eqnarray}
\label{spinpath}
% & = & \frac{1}{2\pi i}\:\int dzdz^{*}\int d\phi^{*}d\phi\, e^{-\phi^{*}\phi}\frac{2S+1}{2\pi(1+|z|^{2})^{2}}<z\phi|e^{-\epsilon H}\ldots e^{-\epsilon H}\ldots e^{-\epsilon H}|z\phi>\\
\mathcal{Z} & = & \tilde{A}\prod_{i}^{M}\int dz_{i}dz_{i}^{*}\int d\phi_{i}^{*}d\phi_{i}\frac{e^{-\phi_{i}^{*}\phi_{i}}}
      {(1+|z_{i}|^{2})^{2}} \langle z_{M}\phi_{M}|e^{-\epsilon H}|z_{1}\phi_{1} \rangle \\
&\times& \langle z_{1}\phi_{1}|\ldots|z_{i}\phi_{i} \rangle
 \langle z_{i}\phi_{i}|e^{-\epsilon H}|z_{i+1}\phi_{i+1} \rangle \nonumber \\
 & \times &\langle z_{i+1}\phi_{i+1}|\ldots|z_{M-1}\phi_{M-1} \rangle
\langle z_{M-1}\phi_{M-1}|e^{-\epsilon H}|z_{M}\phi_{M} \rangle \nonumber,
\end{eqnarray}
where $\tilde{A}=\Big(\frac{2S+1}{4\pi^2 i}\Big)^M$ and $\epsilon=\beta/M$, with $M\gg \beta$.
Each of the matrix elements $\langle z_{i}\phi_{i}|e^{-\epsilon H}|z_{i+1}\phi_{i+1} \rangle$
can then be cast into
\begin{eqnarray}
\langle z_{i}\phi_{i}|e^{-\epsilon H}|z_{i+1}\phi_{i+1} \rangle & = & \langle z_{i}\phi_{i}|1-\epsilon H+O(\epsilon^{2})|z_{i+1}\phi_{i+1} \rangle 
\nonumber \\
 & = & \langle z_{i}\phi_{i}|1-\epsilon[\Gamma\sigma^{x}+\omega a^{\dagger}a+g\sigma^{z}(a^{\dagger}+a)]|z_{i+1}\phi_{i+1} \rangle \nonumber \\
 & = & \langle \phi_{i}|\phi_{i+1} \rangle \langle z_{i}|z_{i+1}\rangle \nonumber \\
&\times&(1-\epsilon[\Gamma \langle z_{i}|\sigma^{x}|z_{i+1} \rangle / \langle z_{i}|z_{i+1} \rangle +\omega\phi_{i}^{*}\phi_{i+1}\nonumber \\
 &  & +g \langle z_{i}|\sigma^{z}|z_{i+1} \rangle /\langle z_{i}|z_{i+1}\rangle (\phi_{i}^{*}+\phi_{i+1})]).
\label{shorttime2}
\end{eqnarray}
Note the difference between the short-time propagators in Eqn.~(\ref{shorttime1}) and (\ref{shorttime2}). 
This is a consequence of the overcompleteness of the $|z \rangle $ basis
versus the orthogonality of the $|\sigma \rangle $ basis.
From Eqs.~(\ref{MEJx})-(\ref{MEJz}), the matrix elements of the spin operators between coherent states
become
\begin{eqnarray*}
\frac{\langle z_{i}|\sigma^{x}|z_{i+1} \rangle}{\langle z_{i}|z_{i+1}\rangle} & = & S\frac{z_{i}^{*}+z_{i+1}}{1+z_{i}^{*}z_{i+1}},\\
\frac{\langle z_{i}|\sigma^{z}|z_{i+1} \rangle}{\langle z_{i}|z_{i+1}\rangle} & = & S\frac{z_{i}^{*}z_{i+1}-1}{1+z_{i}^{*}z_{i+1}},
\end{eqnarray*}
and together with Eq.~(\ref{overlap}), the short-time matrix element
can be rewritten as
\begin{eqnarray*}
\langle z_{i}\phi_{i}|e^{-\epsilon H}|z_{i+1}\phi_{i+1} \rangle & = & \frac{(1+z_{i}^{*}z_{i+1})^{2S}}{(1+|z_{i}|^{2})^{S}(1+|z_{i+1}|^{2})^{S}}\, 
e^{\phi_{i}^{*}\phi_{i+1}}\,\Big(1-\epsilon[\Gamma S \frac{z_{i}^{*}+z_{i+1}}{1+z_{i}^{*}z_{i+1}}\\
 &  & +\omega\phi_{i}^{*}\phi_{i+1}+g S\frac{z_{i}^{*}z_{i+1}-1}{1+z_{i}^{*}z_{i+1}} (\phi_{i}^{*}+\phi_{i+1})]\Big).
\end{eqnarray*}
According to the rules of thumb of path integration, it is only necessary to keep terms of order $\epsilon$ in the short time matrix elements.
If one considers paths that obey $z_{i+1}-z_{i}=O(\epsilon)$ 
one obtains
\begin{eqnarray*}
\frac{\langle z_{i}|\sigma^{x}|z_{i+1}\rangle}{\langle z_{i}|z_{i+1}\rangle} & = & S\:\frac{2\mbox{Re} (z_{i})}{1+|z_{i}|^{2}}\equiv S\, n_{x}^{i},\\
\frac{\langle z_{i}|\sigma^{z}|z_{i+1}\rangle}{\langle z_{i}|z_{i+1}\rangle} & = & S\:\frac{|z_{i}|^{2}-1}{1+|z_{i}|^{2}}\equiv S\, n_{z}^{i}.
\end{eqnarray*}
%Here, we used that the rule of thumb of path integration that we need
%to keep only terms of order $\epsilon$. 
%So,
%\begin{eqnarray}
%e^{-\phi_{i}^{*}\phi_{i}}<z_{i}\phi_{i}|e^{\epsilon H}|z_{i+1}\phi_{i+1}> & = & \frac{(1+z_{i}^{*}z_{i+1})^{2S}}{(1+|z_{i}|^{2})^{S}(1+|z_{i+1}|^{2})^{S}}\, e^{\phi_{i}^{*}\phi_{i+1}-\phi_{i}^{*}\phi_{i}}\,(1-\epsilon[\Gamma\, n_{x}^{i}\nonumber\\
% & + & \omega\,\phi_{i}^{*}\phi_{i+1}+g\, n_{z}^{i}\,(\phi_{i}^{*}+\phi_{i+1})]).\\
% & = & \frac{(1+z_{i}^{*}z_{i+1})^{2S}}{(1+|z_{i}|^{2})^{S}(1+|z_{i+1}|^{2})^{S}}\, e^{\phi_{i}^{*}\phi_{i+1}-\phi_{i}^{*}\phi_{i}}\nonumber \\
%&\times&[e^{\epsilon(\Gamma n_{x}^{i}+gn_{z}^{i}(\phi_{i}^{*}+\phi_{i+1})+\omega\phi_{i}^{*}\phi_{i+1})}+O(\epsilon^{2})]
%\end{eqnarray}
The overlap between spin coherent states for $z_{i+1}-z_{i}\equiv x=O(\epsilon)$
can be cast into\cite{Kuratsuji.80}
\begin{eqnarray}
\langle z_{i}|z_{i+1}\rangle & = & \frac{(1+z_{i}^{*}z_{i+1})^{2S}}{(1+|z_{i}|^{2})^{S}(1+|z_{i+1}|^{2})^{S}}\\
 & \doteq & \frac{(1+|z_{i}|^{2}-z_{i}^{*}x)^{2S}}{(1+|z_{i}|^{2})^{S}(1+|z_{i}|^{2}-x^{*}z_{i}-xz_{i}^{*})^{S}}\nonumber \\
 & = & \frac{(1-\frac{z_{i}^{*}x}{1+|z_{i}|^{2}})^{2S}}{(1-\frac{x^{*}z_{i}+xz_{i}^{*}}{1+|z_{i}|^{2}})^{S}}\nonumber \\
 & \doteq & 1-2S\,\frac{z_{i}^{*}x}{1+|z_{i}|^{2}}+S\frac{x^{*}z_{i}+xz_{i}^{*}}{1+|z_{i}|^{2}}\nonumber \\
 & = & 1+S\,\frac{x^{*}z_{i}-xz_{i}^{*}}{1+|z_{i}|^{2}}\nonumber \\
 & \doteq & \exp \left (S\,\frac{x^{*}z_{i}-xz_{i}^{*}}{1+|z_{i}|^{2}} \right ),
\end{eqnarray}
where the $\doteq$ implies equality up to order $z_{i+1}-z_{i}=x=O(\epsilon)$.
This suggests, that the overlap $\langle z_{i}|z_{i+1}\rangle$ can be related to the image of $z_i$ on $S^2$, provided $z_i$ and $z_{i+1}$ are 
close enough to each other ($\epsilon \rightarrow 0$). The short-time matrix element together with the normalizing factor $\exp[-\phi_{}^{*}\phi_{}]$ 
for the bosonic coherent state becomes
\begin{eqnarray}
e^{-\phi_{i}^{*}\phi_{i}^{}}\langle z_{i}\phi_{i}|e^{-\epsilon H}|z_{i+1}\phi_{i+1}\rangle
 & = & \exp \left (S\,\frac{(z_{i+1}^{*}-z_{i}^{*})z_{i}-(z_{i+1}-z_{i})z_{i}^{*}}{1+|z_{i}|^{2}}\right ) \nonumber \\
 &\times & \exp \left( \phi_{i}^{*}(\phi_{i+1}-\phi_{i}) \right ) \\
 &\times & \exp \left( -\epsilon[\Gamma S n_{x}^{i}+g S n_{z}^{i}(\phi_{i}^{*}+\phi_{i+1})+\omega\phi_{i}^{*}\phi_{i+1}]\right )\nonumber \\
 & = & \exp \left(S\,\epsilon\frac{\frac{(z_{i+1}^{*}-z_{i}^{*})}{\epsilon}z_{i}-\frac{(z_{i+1}-z_{i})}{\epsilon}z_{i}^{*}}{1+|z_{i}|^{2}} \right )\nonumber \\
&\times& \exp \left(\epsilon\frac{\phi_{i}^{*}(\phi_{i+1}-\phi_{i})}{\epsilon} \right)\nonumber \\
&\times&  \exp \left(-\epsilon[\Gamma S n_{x}^{i}+g S n_{z}^{i}(\phi_{i}^{*}+\phi_{i+1})+\omega\phi_{i}^{*}\phi_{i+1}] \right) \nonumber.
\end{eqnarray}
In the limit $M\rightarrow \infty$, while keeping $\beta$ fixed, i.e. $\epsilon \rightarrow 0$, one has
\begin{eqnarray}
\label{phasefac}
e^{-\phi_{i}^{*}\phi_{i}^{}}\langle z_{i}\phi_{i}|e^{-\epsilon H}|z_{i+1}\phi_{i+1} \rangle \Big|_{\epsilon \rightarrow 0} & = &
e^{\epsilon \phi_{i}^{*}\partial_{\tau}\phi_{i}^{}} 
\exp \left( S\,\epsilon\frac{z_{i}^{}\partial_\tau z_{i}^{*}-z_{i}^{*}\partial_\tau z_{i}^{ }}{1+|z_i|^2} \right) \\
& & \hspace*{-1cm}\times \exp \left(-\epsilon[\Gamma S n_{x}^{i}+g S n_{z}^{i}(\phi_{i}^{*}+\phi_{i+1}^{})+\omega\phi_{i}^{*}\phi_{i+1}^{}] \right),\nonumber 
%& = &
% e^{\epsilon\,i S \frac{2 \mbox{\tiny Im}(z_{i}^{*}\partial_\tau z_{i}^{})}{1+|z_i|^2}}
%e^{\epsilon \phi_{i}^{*}\partial_{\tau}\phi_{i}}\nonumber \\
%&\times&  e^{\epsilon[\Gamma S n_{x}^{i}+g S n_{z}^{i}(\phi_{i}^{*}+\phi_{i+1})+\omega\phi_{i}^{*}\phi_{i+1}]}\nonumber 
\end{eqnarray}
where 
$\lim_{\epsilon\rightarrow 0}\frac{\phi_{i}^{*}(\phi_{i+1}-\phi_{i})}{\epsilon}=\phi^*\frac{\partial \phi}{\partial \tau}\equiv \phi^*\partial_{\tau}\phi$
and likewise for $(z_{i+1}-z_{i})/\epsilon$ was used. 
Note, 
$\epsilon\,i S\, \frac{2\mbox{\small Im}(z_{i}^{*}\partial_{\tau} z_{i}^{})}{1+|z_i|^2}$ is purely imaginary, so that  the first term on the right hand side of Eq.~(\ref{phasefac}) becomes a pure phase factor.
Since $\vec{n}^2(\tau)=n_x^2(\tau)+n_y^2(\tau)+n_z^2(\tau)=1$, the time derivative is tangential to $S^2$ making 
$\frac{z_{i}^{}\partial_\tau z_{i}^{*}-z_{i}^{*}\partial_\tau z_{i}^{ }}{1+|z_i|^2}$ a differential one-form $\alpha$ on $S^2$.  
Inserting the short-time matrix element, Eq.~(\ref{phasefac}), back into Eq.~(\ref{spinpath}), the continuum limit poses no further 
difficulties, since we are dealing with c-numbers. The pure phase factors along the path of the system add up and according to
Stokes theorem are equivalent to the area on
the sphere $S^2$ traced out by $\vec{n}(\tau)$ ($0\leq \tau\leq \beta$ with $\vec{n}(0)=\vec{n}(\beta)$),  ${\mathcal A}[\vec{n}(\tau)]$:
\begin{equation}
i S \int_{\partial {\mathcal A}} \alpha = i S \int_{\mathcal{A}} d\alpha\equiv i S {\mathcal A}[\vec{n}].
\end{equation}
Figure \ref{berryphase} shows the area ${\mathcal A}[\vec{n}(\tau)]$  associated with a particular path $\vec{n}(\tau)$.

The path integral representation for the partition function of Eq.~(\ref{Hamiltonian}) finally becomes
\begin{eqnarray}
{\mathcal{Z}}&=&\int{\mathcal{D}}[\vec{n},\phi^*,\phi]\exp \Big[ i S {\mathcal A}[\vec{n}]\Big]\\
&\times& \exp \Big[
\int_0^{\beta}d\tau \big(\phi^{*}\partial_{\tau}\phi^{} - \Gamma S n_x(\tau)+ g S n_z(\tau)(\phi^{*}(\tau)+\phi^{}(\tau))+\omega\phi^{*}(\tau)\phi^{}(\tau) \big)\Big]. \nonumber
\end{eqnarray}
%%%%%%%%%%%%%%%%%%% Figure 3 %%%%%%%%%%%%%%%%%%%%%%%%%%%%%%%%%%%%%%%%%%%%%%%%%%%%%%%%%%%%%%%%%
%%%%%%%%%%%%%%%%%%%%%%%%%%%%%%%%%%%%%%%%%%%%%%%%%%%%%%%%%%%%%%%%%%%%%%%%%%%%%%%%%%%%%%%%%%%%%%
The partition function  is now in a form that permits to globally integrate out the bosonic mode via Eq.~(\ref{integraltransform}).\\

For the sub-Ohmic spin-boson model, Eq.~(\ref{spinboson}), the partition function for arbitrary spin $S$ therefore becomes
\begin{eqnarray}
{\mathcal{Z}}&=&\int{\mathcal{D}}[\vec{n}]\exp[-S_{\mbox{\tiny eff}}],
\label{eq:SBpartition}
\end{eqnarray}
and the effective action for the spin degrees of freedom is given by
 \begin{eqnarray}
\label{Seff}
S_{\mbox{\tiny eff}}&=& -i S  {\mathcal A}[\vec{n}]+\!\,\! \int_0^\beta d\tau \Gamma S n_x(\tau)+
g^2S^2 \int_0^\beta d\tau \int_0^{\beta}d \tau^{'} n_z(\tau)\chi^{-1}_0(\tau-\tau^{'})n_z(\tau^{'}),~~~~~~~
\end{eqnarray}
where\cite{Kirchner.09}
\begin{equation}
\label{chiInv}
\chi^{-1}_0(\tau-\tau^{'})=\int^{\infty}_{0}d\omega \omega^{1-\eta}\frac{\cosh[\omega(\beta/2-|\tau-\tau^{'}|)]}{\sinh(\omega\beta/2)},
\end{equation}
from Eq.~(\ref{spectraldensity}) with $\Lambda \rightarrow \infty$.

This is the central result: the effective action of the sub-Ohmic spin-boson problem in the continuum limit is given by
Eqs.~(\ref{eq:SBpartition})-(\ref{chiInv}) and contains a Berry phase term. An analysis along the lines presented in section \ref{sec:GLW}
can therefore  not readily be applied. That a Berry phase term in the effective action can change critical properties is known e.g. from 
the spin-isotropic SU(N) BFKM in the limit of large N, where the model displays critical exponents that differ from those of its classical 
counterpart~\cite{Kirchner.08e}. Whether the quantum-to-classical mapping fails for the sub-Ohmic spin-boson problem or not is at present unclear.
The reduced symmetry of the spin-boson Hamiltonian (as compared to the spin-isotropic BFKM) makes the analysis of the critical properties much harder.
Still, the Berry phase term in the path integral implies the importance of topological excitations~\cite{Kirchner.09}.

\section{Conclusions}

In this article, I showed that the effective action in the continuum limit  for the spin degrees of freedom in the sub-Ohmic spin-boson model 
with arbitrary spin $S$ necessarily contains
a Berry phase term which results from the spin-coherent state basis. This is a consequence of the coupling to bosons and their continuous paths
in the bosonic phase space.  
In a spin-only problem, a functional integral based on an orthogonal basis can be constructed~\cite{Fahri.92,Krzakala.08}. For a single quantum 
spin in a transverse field, the resulting action is equivalent to that of a classical Ising spin chain, as anticipated.   
When the quantum spin is coupled to the bosonic bath, it turns out that the continuum limit ($M\rightarrow \infty$ in Eq.~(\ref{partition1})) 
in the eigenbasis of $\sigma^z$, i.e. the Ising degrees of freedom,  
cannot be taken in a straightforward manner and the quantum-to-classical mapping to a classical Ising chain with long-ranged interaction
cannot easily be carried through. The problems that underlie the continuum limit can be traced back to the spin flips from one eigenvalue of $\sigma^z$ to
the other.
A physical interpretation of why the limit of vanishing lattice constant (while number of Ising spins approaches infinity)
of a classical Ising chain with a long-ranged interaction
(corresponding to the retarded interaction generated by the bosons) fails to be equivalent to the sub-Ohmic Bose-Fermi Kondo model, was given in 
reference~\cite{Kirchner.09}. This line of arguments is therefore complementary to the approach taken here.
Finally, I demonstrated that a  spin-coherent state basis circumvents the associated difficulties, so that both the continuum limit can be performed
and the bosonic field can be integrated out.  To this end, a basic introduction to spin coherent states was presented that draws from the analogy
to bosonic coherent states. The functional integral for spin in terms of spin coherent states is formulated on the unit sphere $S^2$, the coset space of
SU(2). 
Since the SU(2) group can decomposed as SU(2)=$S^2\times$U(1) and U(1) is multiply connected while SU(2) is simply connected, any coordinate system on 
$S^2$ must necessarily be singular. As a result, a Berry phase term arises in the spin path integral out of this geometric property  of $S^2$ and
the effective action for the spin degrees of freedom does not assume the form of a 
Ginzburg-Landau-Wilson functional. That such a Berry phase is responsible for the non-classical behavior of the totally spin-isotropic, sub-Ohmic
 Bose-Fermi Kondo 
model has been demonstrated in reference~\cite{Kirchner.08e}. Numerical renormalization group (NRG)
 studies on the easy-axis Bose-Fermi Kondo model, which is closely related to the
spin-boson model discussed here, indeed suggest a breakdown of the quantum-to-classical mapping for this model as well~\cite{Glossop.05,Glossop.07b}. 
The applicability of the NRG
to critical systems involving bosons has recently been called into question~\cite{Vojta.10,Vojta.09}. The main supportive evidence 
for such an  apparent shortcoming of the NRG method seems to have come from a recent  classical
Monte Carlo study~\cite{Winter.09}, 
whose relevance to the spin-boson problem, as I have shown here, needs to be re-examined.

\begin{acknowledgements}
I thank Carlos Bolech, Kevin Ingersent, Akira Inomata, 
Steffen Wirth and in particular Qimiao Si for many helpful discussions. A fruitful collaboration with Kevin Ingersent and
Qimiao Si on issues related to this article is greatly acknowledged and appreciated. 
\end{acknowledgements}

%\bibliographystyle{spmpsci.bst}
%\bibliography{QPT2010}

\end{document}